\documentclass[useAMS,usenatbib,aas_macros,twocolumn]{mn2e}

\usepackage{epsfig}
\usepackage{epsf}
\usepackage{lscape,graphicx,colortbl}
\usepackage{natbib}
\usepackage{times}
\usepackage{aas_macros}

\usepackage{natbib}
\usepackage{amsmath,amssymb}
\usepackage{mathptmx}
\usepackage{graphicx}
\usepackage{longtable}
\usepackage[font=small,labelfont=bf]{caption}
\usepackage{graphicx,rotating,booktabs}
\usepackage[verbose]{placeins}
\usepackage{blindtext}
\usepackage{graphicx,rotating,booktabs}
\usepackage{amsmath}
\newcommand  \ergs     {\ifmmode {\rm erg\,s}^{-1} \else erg s$^{-1}$\fi}
\newcommand  \Msunyr     {\ifmmode {\rm \Msun\,yr}^{-1} \else \Msun\ yr$^{-1}$\fi}
\newcommand  \msunyr     {\ifmmode {\rm \Msun\,yr}^{-1} \else \Msun\ yr$^{-1}$\fi}
\newcommand  \cmii     {\ifmmode {\rm cm}^{-2} \else cm$^{-2}$\fi}
\newcommand  \cmiii     {\ifmmode {\rm cm}^{-3} \else cm$^{-3}$\fi}
\def\Hubble{\ifmmode {\rm km\,s}^{-1}\,{\rm Mpc}^{-1}\else km\,s$^{-1}$\,Mpc$^{-1}$\fi}
\def\Msun{\ifmmode M_{\odot} \else $M_{\odot}$\fi}
\def\msun{\ifmmode M_{\odot} \else $M_{\odot}$\fi}
\def\Lsun{\ifmmode L_{\odot} \else $L_{\odot}$\fi}
\def\Zsun{\ifmmode Z_{\odot} \else $Z_{\odot}$\fi}
\def\qo{\ifmmode q_{0} \else $q_{0}$\fi}
\def\Ho{\ifmmode H_{0} \else $H_{0}$\fi}
\def\ho{\ifmmode h_{0} \else $h_{0}$\fi}
\def\qo{\ifmmode q_{0} \else $q_{0}$\fi}
\def\ao{\ifmmode a_{0} \else $a_{0}$\fi}
\def\to{\ifmmode t_{0} \else $t_{0}$\fi}
\def\omm{\ifmmode \Omega_{{\rm M}} \else $\Omega_{{\rm M}}$\fi}
\def\omlam{\ifmmode \Omega_{\Lambda} \else $\Omega_{\Lambda}$\fi}

\def\mgii{\ifmmode {\rm Mg}{\textsc{ii}} \else Mg\,{\sc ii}\fi}
\newcommand \MgII {\ifmmode {\rm Mg}\,{\sc ii}\,\lambda2798 \else Mg\,{\sc ii}\,$\lambda2798$\fi}
\def\Hbeta{\ifmmode {\rm H}\beta \else H$\beta$\fi}

\def \L3000a{$L_{3000}$}
\def \L1450{$L_{1450}$}   
\def \Cf {$C_f$}
\def \L210{$L_{2-10}$}
\newcommand{\kbol}  {\ifmmode k_{\rm BOL} \else $k_{\rm BOL}$\fi}
\newcommand{\lbol}  {\ifmmode L_{\rm bol} \else $L_{\rm bol}$\fi}
\newcommand{\Lbol}  {\ifmmode L_{\rm bol} \else $L_{\rm bol}$\fi}
\newcommand{\lagn}  {\ifmmode L_{\rm AGN} \else $L_{\rm AGN}$\fi}
\newcommand{\LAGN}  {\ifmmode L_{\rm AGN} \else $L_{\rm AGN}$\fi}
\newcommand{\LX}    {\ifmmode L_{\rm X} \else $L_{\rm X}$\fi}
\newcommand{\Lion}  {\ifmmode L_{\rm ION} \else $L_{\rm ION}$\fi}
\newcommand{\LION}  {\ifmmode L_{\rm ION} \else $L_{\rm ION}$\fi}
\newcommand{\LUV}   {\ifmmode L_{\rm 1350} \else $L_{\rm 1350}$\fi}
\newcommand{\Ldust}   {\ifmmode L_{5\mu m} \else $L_{5 \mu m}$\fi}
\newcommand{\lsf}   {\ifmmode L_{\rm SF} \else $L_{\rm SF}$\fi}
\newcommand{\LSF}   {\ifmmode L_{\rm SF} \else $L_{\rm SF}$\fi}
\newcommand{\LTOR}   {\ifmmode L_{\rm torus} \else $L_{\rm torus}$\fi}
\newcommand{\LIR}   {\ifmmode L_{\rm IR} \else $L_{\rm IR}$\fi}
\newcommand{\LHD}   {\ifmmode L_{\rm HD} \else $L_{\rm HD}$\fi}
\newcommand{\lledd} {\ifmmode L/L_{\rm Edd} \else $L/L_{\rm Edd}$\fi}
\newcommand{\Ledd} {\ifmmode L/L_{\rm Edd} \else $L/L_{\rm Edd}$\fi}
\newcommand{\LEDD} {\ifmmode L/L_{\rm Edd} \else $L/L_{\rm Edd}$\fi}
\newcommand{\fwmg}  {\ifmmode {\rm FWHM}\left(\mgii\right) \else FWHM(\mgii)\fi}
\newcommand{\CFHD}  {\ifmmode {\rm CF}_{\rm HD} \else ${\rm CF}_{\rm HD}$\fi}
\newcommand{\mbh}   {\ifmmode M_{\rm BH} \else $M_{\rm BH}$\fi}
\newcommand{\MBH}   {\ifmmode M_{\rm BH} \else $M_{\rm BH}$\fi}
\newcommand{\mstar}   {\ifmmode M_{*} \else $M_{*}$\fi}
\newcommand{\Mstar}   {\ifmmode M_{*} \else $M_{*}$\fi}
\newcommand{\Mdot}{\ifmmode \dot{M} \else $\dot{M}$\fi}
\newcommand{\mdot}{\ifmmode \dot{m} \else $\dot{m}$\fi}

\def  \MgII         {\ifmmode {\rm Mg}\,{\sc ii}\,\lambda2798
                  \else Mg\,{\sc ii}\,$\lambda2798$\fi}
\def  \mgii         {\ifmmode {\rm Mg}\,{\sc ii} \else Mg\,{\sc ii}\fi}
\def  \OIII         {\ifmmode {\rm [O}\,{\sc iii]}\,\lambda5007
                  \else [O\,{\sc iii]}\,$\lambda5007$\fi}
\def  \oiii         {\ifmmode {\rm [O}\,{\sc iii]}\,\lambda5007
                  \else [O\,{\sc iii]}\,$\lambda5007$\fi}
\def  \oi         {\ifmmode {\rm [O}\,{\sc i]}\,\lambda6300
                  \else [O\,{\sc i]}\,$\lambda6300$\fi}
\def\ha{\ifmmode {\rm H}\alpha \else H$\alpha$\fi}
\def\Ha{\ifmmode {\rm H}\alpha \else H$\alpha$\fi}
\def\hb{\ifmmode {\rm H}\beta \else H$\beta$\fi}
\def\Hb{\ifmmode {\rm H}\beta \else H$\beta$\fi}
\def\La{\ifmmode {\rm L}\alpha \else L$\alpha$\fi}

\def\Chisq{\ifmmode \chi^{2} \else $\chi^{2}$}
\def \zzz {\ifmmode z=2-3.5 \else $z =2-3.5$\fi}

\def\z48{$z \simeq$4.8}
\def\z33{$z \simeq$3.3}
\def\z24{$z \simeq$2.4}

\title[Bolomtric correction factors for AGN]
{Bolometric correction factors for Active Galactic Nuclei}

\author[Hagai Netzer]
{Hagai Netzer $^1$\thanks{E-mail: hagainetzer@gmail.com} \\
$^1$School of Physics and Astronomy, Tel-Aviv University, Tel-Aviv 69978, Israel
}

\begin{document}
\maketitle




\begin{abstract}

The bolometric luminosity of active galactic nuclei (AGN)
is difficult to determine and various approximations 
have been used to calibrate it against different observed properties.
Here I combine theoretical calculations of optically thick, geometrically thin accretion disks, and 
observed X-ray properties of AGN,  
to provide new bolometric correction factors (\kbol) over a large range of
black hole (BH) mass, accretion rate, and spin. 
 This is particularly important in cases where the mass accretion rate cannot
be determined from the observed spectral energy distribution, and in cases where luminosity-independent correction factors
have been used.  
Simple powerlaw approximations of \kbol\ are provided for L(5100\AA), L(3000\AA), L(1400\AA), L(2-10 keV) and L(narrow \hb). 
In all cases 
 the uncertainties are large mostly due to the unknown BH spin.
Prior knowledge of the BH mass reduces the uncertainty considerably.

\end{abstract}

\begin{keywords}

(galaxies:) quasars: general; galaxies: nuclei; galaxies: active; accretion, accretion discs
\end{keywords}

\section{Background}

A major challenge in the study of active galactic nuclei (AGN) is the estimate of their bolometric luminosity (hereafter \LAGN).
The main limitation is that much of the rest-frame ultraviolet (UV) spectral energy distribution (SED) is hidden from view, either due to galactic absorption in low 
redshift objects, or inter-galactic absorption in high redshift
objects. Additional difficulties are the limited wavelength range accessible to various survey like SDSS, WISE and UKIDSS, the variability which is hard to measure, and dust extinction in some objects.
The uncertain \LAGN\ affects the estimate of black hole (BH) growth and the Eddington ratio (\Ledd). It also limits our understand of  
 the interaction of the central radiation source
with the central accretion disk (AD), the broad line region (BLR), the narrow line region (NLR), the central dusty torus,
the interstellar matter (ISM) in the host galaxy, and the inter-galactic medium at larger distances. These issues have been discussed in numerous 
publications \cite[for recent reviews see][]{Beckmann2012,Netzer2013}.

AGN are probably powered by the conversion of gravitational energy to electromagnetic radiation via various types of 
accretion flows. Examples of such flows are
radiatively efficient, optically thick, geometrically
thin ADs, \cite[e.g][hereafter SS73]{Shakura1973}, radiatively inefficient, 
advection dominated accretion flows \cite[ADAF, see e.g.][]{Narayan2005}, 
and very high accretion rate, geometrically thick or slim disks
 \cite[e.g.][]{Mineshige2000,Sadowski2016}.
Radiatively efficient, geometrically thin ADs are perhaps the best understood accretion flows. Their total radiative power is determined by the BH accretion rate (\Mdot) 
and spin (represented by the spin parameter {\it a}), and their theoretical  
SEDs have been compared, successfully, with  numerous observations.
The mass to radiation conversion efficiency in such disks, $\eta$, ranges between 0.321 ($a=0.998$) and 0.038 ($a=-1$). 
Perhaps the most detailed observations of such disks, in terms of rest-frame wavelength coverage, are those presented by 
\cite{Capellupo2015,Capellupo2016}. They show a very good agreement between the predicted thin AD spectra and the observations.
Various other publications, 
based on a more restricted wavelength range \cite[see][]{Davis2007,Jin2012a,Jin2012b,Lusso2015}, reach similar conclusions.

The higher accretion rate slim ADs are common but not so well understood. Such systems are identified either by their high normalized
accretion rate \Mdot/\MBH, or by their unique spectroscopic properties. 
Main candidates to host such disks are Narrow Line Seyfert 1 galaxies (NLS1s) or high redshift very luminous AGN, both with
\Ledd$\sim 1$. There are attempts to calculate the SED
of such sources  
using thin AD models with modified local temperature and opacity 
 and various types of coronae
\cite[see e.g.][]{Done2012,Done2013,Kubota2018}.  
Other models invoke saturated disk radiation above a certain \Ledd, 
due to radial advection, and/or powerful disk winds which are capable  
of carrying out a significant part of
the released gravitational energy. Under these conditions, the connection between accretion rate, spin
 and radiated power, can differ substantially  
from those in thin ADs.
The few available numerical calculations \cite[e.g.][]{Sadowski2016}, suggest overall
radiative efficiency which is lower than the efficiency of thin ADs, perhaps in the range 0.01-0.08. 

The low efficiency flows, 
like ADAFs, are also less well understood.  Here, again, 
there is no simple relationship between \LAGN, BH spin, and accretion
rates, 
 and 
 $\eta$ is low, perhaps below $10^{-3}$. 
The present work focus on geometrically thin, optically thick ADs; the main power-house of most observed high ionization AGN, 
but perhaps not all LINERs.

Most AGN are powerful X-ray emitters with energy in the approximate range 0.2-100 keV,
where the limit at 0.2 keV is arbitrarily determined by the capability
of most X-ray instruments.
 In radio-quiet AGN, the emitted X-ray radiation is though to be drawn, entirely, from the 
gravitational energy of the accretion
flow. The origin of the X-ray radiation is still somewhat unclear. The soft X-ray radiation 
may be associated with a Compton thick medium at the surface of the disk \cite[e.g.][and references therein]{Done2012}. At higher energies, 
the emitted radiation is probably produced 
by a Compton thin  corona which extends over the central parts of the disk \cite[see e.g.][]{Done2012,Reynolds2016,Kubota2018}. 
While not all the details of these processes are understood, the relations between the total X-ray luminosity \LX, (in this
work the integrated 0.2-100 keV luminosity) and \LAGN, are well constrained, because these parts of the SED are readily accessible for X-ray instruments.
In this paper I use the 2-10 keV energy, L(2-10 keV), as a proxy for \LX.

Some of the emitted disk radiation can ionize low-density gas in the NLR
 which results in easy to observe narrow emission lines. The lines can be used to estimate \LAGN\ provided 
the gas is optically thick to Lyman continuum radiation, and its covering factor (\Cf), and amount of dust extinction, are known. Detailed investigation of these
issue have been published by \cite{Heckman2004}, \cite{Netzer2009b}, \cite{Heckman2014}, 
and others. 

The purpose of this paper is to calculate various optical, UV and X-ray bolometric correction factors (hereafter \kbol) 
for AGN that are powered by optically thick, geometrically thin ADs 
 that extend all the way to the innermost stable circular orbit (ISCO).
It follows  several earlier publications discussing this issue
\cite[e.g.][]{Marconi2004,Vasudevan2007,Richards2006,Netzer2009b,Nemmen2010,Runnoe2012a,Jin2012a,Trakhtenbrot2012,Krawczyk2013,Netzer2016}.
Improving the accuracy of such estimations is required in many cases where the accretion rate cannot be determined from the optical SED either
because of the redshift or the very high luminosity of the  sources \cite[see e.g. Fig. 1 in][]{Netzer2014a}. 

 In \S2 I describe the accretion disk and X-ray models and explore \kbol\ over a large range of BH mass, spin, and accretion rate.
In \S3 I discuss the implications to AGN study and in \S4 I summarize the results.
 
\section{Calculations}
This work provides theoretically-based calculations for a variety of bolometric correction factors:
\kbol(5100\AA), \kbol(3000\AA), \kbol(1400\AA), \kbol(2-10 keV) and 
\kbol(narrow \hb), where \LAGN=$\kbol \times (observed\,luminosity)$ and L(1400\AA) stands
for  $\lambda L_{\lambda}$ at 1400\AA.  
The calculations are limited to optically thick, geometrically thin ADs assuming that accretion through the disk, 
 with time independent accretion rate, is the only energy production mechanism.
and \LAGN=$\eta \Mdot c^2$.  For reasons explained below, I avoid slim accretion disks with \Ledd$>0.5$. I also avoid very low accretion rate systems, with \Ledd$<0.001$. 
The emitted X-ray radiation is assumed to be drawn from the same source of energy, thus
 the calculated \LX\ does not affect the derived \LAGN.
The range of BH mass is $10^7-10^{10}$~\msun,  and the BH spin parameter covers the entire range allowed for such systems, from -1 to 0.998.

\subsection{Accretion disk calculations}

The accretion disk model used here is described in \cite{Slone2012}. It is similar to the standard SS73 disk but includes also full relativistic corrections and
Comptonization in the disk atmosphere. The disk is assumed to be stationary with a constant viscosity parameter $\alpha=0.1$. 
It extends from the ISCO out to $r=2160 r_g$, where $r_g$ is the gravitational radius.
The calculations start with an assumed BH mass, BH spin and disk inclination.  
The entire SED is then calculated for a range of accretion rates such that the entire allowed range of \Ledd\ is covered. 
The resulting spectra are used to define \kbol(1400\AA), \kbol(3000\AA) 
and \kbol(5100\AA).

Several other types of AGN disks have been considered. In particular, two or three-part disks involving warm and hot X-ray gas, have been
suggested \cite[e.g.][and references therein]{Done2012,Kubota2018,Panda2019}. Such systems produce significantly different ionizing SEDs and
are discussed, briefly, in \S2.2. The code {\it QSOSED} is available as a part of
 {\it XSPEC} for calculating such SEDs.

The disk inclination factor assumed in the calculations takes a simple, wavelength-independent form. 
\begin{equation}
\frac{L_{\nu}}{L_{\nu}(face-on)}=\frac{1}{1+b} \cos \theta (1+b \cos \theta)
\end{equation}
where $\theta$ is the inclination to the line of sight and $b$ a limb darkening factor. 
The case $b=0$ corresponds to the simplest case of   $L_{\nu} \propto \cos \theta$.
Here I use $b=2$ which is in good agreement with an electron scattering atmosphere. There are some differences between the two assumed values of $b$ regarding the
dependence of the inferred \kbol\ on the (observationally unknown) disk inclination. For type-I AGN, with $\cos \theta \sim 0.7$, they amount to only a few percent. For very large inclinations,
those hardly observed in type-I sources, they are much larger.
The values of \kbol(1400\AA), \kbol(3000\AA)
 and \kbol(5100\AA) were calculated for a disk inclination that recovers the correct \LAGN\ by integrating 
$L_{\nu}$ over the observed SED ($\theta \approx 56$ degrees).

As shown in \cite{Slone2012}, and also in \cite{Laor2014},
disk winds can reduce \LAGN, soften the emitted spectrum  
and affect \kbol. Such winds are not considered in this work. I also do not consider various line and bound-free absorption properties in the disk atmosphere 
that can significantly 
alter the far-UV SED \cite[see e.g.][]{DavisLaor2011} but not 
the total emitted power. As shown below, earlier calculations of \kbol\ for  geometrically thin ADs, by \cite{Nemmen2010}, resulted in
 bolometric correction factors   similar
to the ones presented here.

GR effects change photon trajectories close to the ISCO. For the BH mass in question, such effects are only 
noticeable at very large disk inclination and  high spin
\cite[see e.g.][]{Laor1990,Campitiello2018}.
Very few, if any type-I AGN are observed at such inclination angles. In addition, the angular radiation pattern 
of the X-ray source can be different from that of the geometrically thin AD.
 This can have important observational consequences \cite[][]{Netzer1987} but is beyond the scope of the present work.

For illustration purpose, I only show four values of the spin parameter: -1, 0, 0.7 and 0.998. The corresponding mass-to-radiation 
conversion deficiencies are 0.038, 0.057, 0.104 and 0.321. I also show only four values of BH mass, 
$10^7$~\Msun, $10^8$~\Msun, $10^9$~\Msun, and $10^{10}$ \Msun.
This combination of BH mass and spin is enough
to illustrate the entire range of \kbol.
I exclude cases where the fractional ionizing luminosity is below 5\%. For smaller fractions, the resulting equivalent width (EW) of the strongest broad emission lines 
are too small to be detected in
big spectroscopic surveys, like SDSS. Such objects would normally not be classifies as type-I AGN 
because of their very weak, small EW broad emission 
lines \cite[see][]{LaorDavis2011,Hryniewicz2010,Netzer2014a,Bertemes2016}.

The transition between thin and slim accretion disks depends on \Mdot/\MBH\ which, for thin ADs, is proportional to \Ledd. The number chosen here, \Ledd=0.5, is somewhat arbitrary
 mostly because slim disk calculations 
are still rather uncertain. Proposed numbers range from \Ledd=0.1 to \Ledd$\approx 1$ \cite[e.g.][and references therein]{Sadowski2011,Du2015,Sadowski2016}.
The transition to ADAF is taken to be \Ledd$=0.001$. This is also model dependent  
and several calculations assume lower values \cite[see][and references therein]{Sadowski2017}. This limit
leaves many LINERs outside the population considered here. As explained below, various other constraints further
reduce the range of \Ledd.

\subsection{X-ray calculations}
The X-ray properties of AGN have been studied, extensively, in numerous papers 
\cite[][and references therein]{Vasudevan2007,Done2012,Done2013,Jin2012b,Reynolds2016,Kubota2018,Panda2019}. 

 X-ray production must be related to gas with a temperature significantly above the standard thin disk temperature. Several
recent studies \cite[e.g.][]{Kubota2018,Panda2019} consider this radiation to originate from the regions
nearest to the BH. These are divided into {\it warm corona}, producing the so-called ``soft X-ray excess'', 
and {\it hot corona} producing the X-ray power-law.
 The fraction of the accretion energy dissipated in the two regions depends
both on the accretion rate and the Eddington ratio.   

 Here I take a purely
observational approach and consider only the hard X-ray  
 power-law sources, with $N_{E} \propto E^{-\Gamma}$, where $N_E$ is the photon flux at energy E  chosen here to extend
 from 0.2 to 100 keV.
In most AGN $1.6<\Gamma<2.2$ and the value itself must be related to the details of the accretion process.
The value adapted here, $\Gamma=1.9$, is representative of the population of radio quite AGN \cite[][and references therein]{Ricci2017}.
This slope results in 
\LX=3.9L(2-10 keV) where \LX=L(0.2-100 keV).

A central assumption of this work is the existence of a tight correlation between the  UV and X-ray luminosities. 
This correlation, or its equivalent, $\alpha_{OX}$,
have been studied in numerous papers \cite[e.g.][and references therein]{Just2007,Jin2012b}. The form used here is adapted from the various fits presented in \cite{Lusso2016} and is given by:
\begin{equation}
 \log L_{\nu (\rm 2\, keV)}= 0.62 \log L_{\nu \, (2500A)} + 7.77 \,\, .
\end{equation}
The measured scatter around this relationship depends on the X-ray slope $\Gamma$,  on the (unknown) disk inclination, and on the way used to model the observed  
flux at
2500\AA\ which is contaminated by broad FeII lines and Balmer continuum emission \cite[see][and references therein]{Mejia-Restrepo2016}.

Two general assumptions related to the X-ray radiation are used: 
1) Full energy conservation (the radiated X-ray energy is drawn from the gravitational energy of
the flow).  
2) The production of X-ray photons does not affect {\it the measured} L(5100\AA), L(3000\AA) and L(1400\AA). 
The second assumption requires some justification.

As explained, the hard X-ray radiation is probably the results of a hot, Compton-thin corona covering the central
part of the disk and up-scatters 
the optical-UV continuum photons. Energetically, this involves heating of the corona, through dissipation,
 and changing the energy of the scattered radiation. 
Unlike the three region disks considered by \cite{Kubota2018}, the disks in this work extend all the way to the ISCO. In this case,
heating the corona can be achieved if a small fraction of the accretion flow,
  goes through the top layer of the disk and heats it to temperatures far above the thin AD temperature.
This dissipation
reduces the optical-UV luminosity compare with a case of a thin AD without a corona. If the corona covers
only the innermost part of the disk, where most of the emitted flux is beyond 1 Rydberg, then the $E<1 $ Rydberg radiation,
including L(5100\AA), L(3000\AA) and L(1400\AA), is not affected much by the entire process. 
Moreover, in this case most of the upscattered photons are Lyman continuum 
photons. Under such conditions, \LX$<$\LION, where \LION\ stands for  the $E>1$ Rydberg luminosity of a disk  with the same accretion rate but
without a corona.
Since low spin, low accretion rate and large BH mass result in small \LION/\LAGN, and \LX\ is assumed to depend directly on L(2500\AA) (eqn. 2), 
 in some of these cases \LX\ can be large even when \LION/\LAGN\ is very small.     
 Under such conditions, the lowest \Ledd\ that can be achieved is about 0.007, rather than the general low limit of 0.001 imposed earlier.

The alternative is a larger corona that extends far beyond the ISCO. In this case, there is a direct link between the
$E<1$ Rydberg radiation  and \LX\ since the flow going through the corona
reduces the emission from the parts of the disks where this radiation is emitted. Moreover, some of these photons are 
converted to X-ray radiation. 
This requires to impose a second condition of 
 \LX$< 0.5$\LAGN.  There are only a few known AGN where \LX\ is larger than the 
second limit. 

All the calculations presented in this work apply to both conditions, \LX$<$\LION\ and \LX$< 0.5$\LAGN.  For 
a detailed discussion of various possible geometries and types of coronae see
\cite{Done2012} and \cite{Kubota2018}.

\onecolumn

\begin{figure}
\centering
\includegraphics[width=19cm]{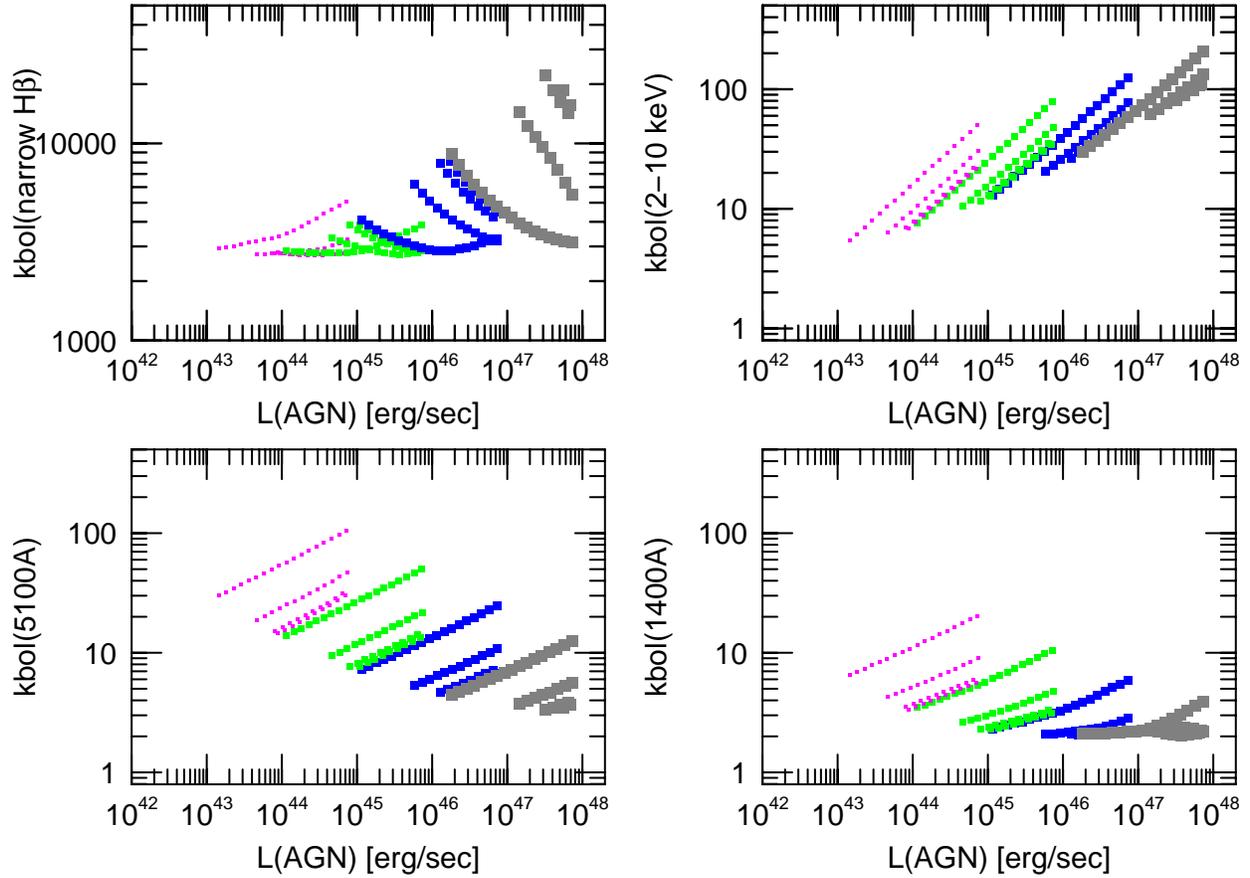}
 \caption{
Various bolometric correction factors vs. \LAGN\ 
for thin accretion disks with ``mean inclination''  and X-ray properties as defined in this work.
Here and in all other diagrams, point size represent BH mass.
 \MBH=$10^7$ \Msun\ (tiny points), \MBH=$10^8$ \Msun\ (small points), \MBH=$10^9$ \Msun\
 (large points) and \MBH=$10^{10}$\Msun\ (largest points).
Each BH mass is separated into four groups of spins: {\it a=-1} (smallest \kbol), {\it a=0, a=0.7} and {\it a=0.998} (largest \kbol).
The top left panel shows \kbol(narrow \hb) assuming optically thick NLR with a covering factor of 0.05 and no internal dust.
}
\label{fig:fig1}
\end{figure}

\begin{figure}
\centering
\includegraphics[width=19cm]{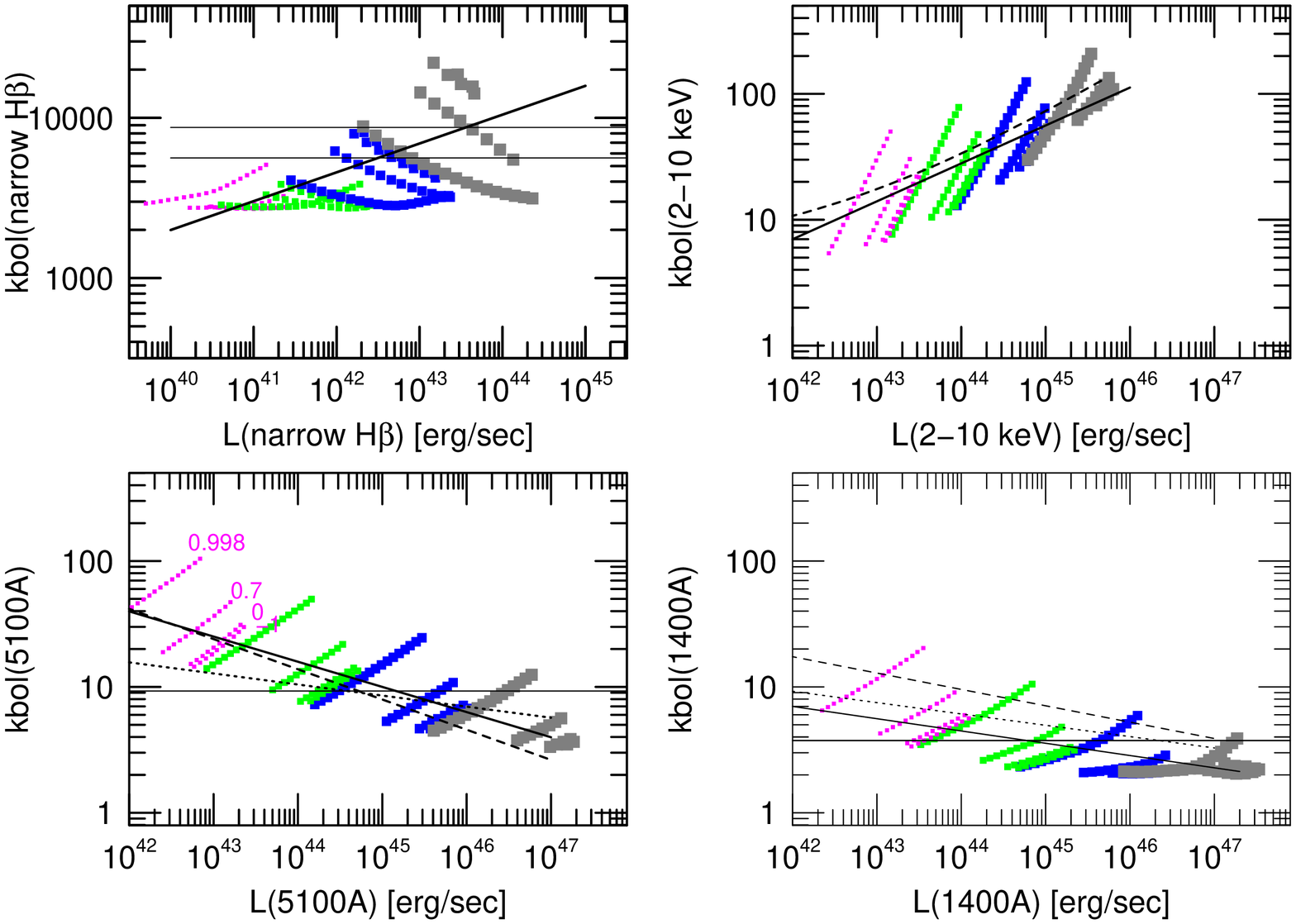}
 \caption{
As Fig.~\ref{fig:fig1} but for observed luminosities, as marked on the horizontal axis.
The spin parameters ({\it a}) for \MBH=$10^7$\Msun\ are marked in the bottom left panel.
The black solid diagonal lines in all panels show the eye-fitted approximations given in eqn. 3 and Table 1.
For \kbol(1400\AA) and \kbol(5100\AA) I also show the AD-based fit of \citet{Nemmen2010} (dashed black lines),
the expression derived for the power-law continuum by \citet{Runnoe2012a} (dotted black lines), and the constants
used in the \citet{Shen2011} catalog following the \citet{Richards2006} analysis (solid black horizontal lines).
The range marked by the two horizontal lines in \kbol(narrow \hb) is the one recommended in \citet{Netzer2009b}.
For \kbol(2-10 keV), the dashed black line is the \citet{Marconi2004} approximation.
}
\label{fig:fig2}
\end{figure}

\twocolumn

\subsection{Narrow emission line calculations}

Several narrow emission lines have also been used to estimate \LAGN. The most useful ones are \ha, \hb, \oiii, and \oi\
 \cite[see][and references therein]{Netzer2009b,Heckman2014,Pennell2017}.
In the present study, the Balmer line luminosities were calculated from the number of ionizing photons, assuming a nominal covering factor
 (\Cf) for the NLR and a 
 standard Case-B recombination with $T_e=10^4$K. Since L(narrow \ha)/L(narrow \hb) is known from recombination
theory, I only show \kbol(narrow \hb). 
Realistic photoionization calculations must also take into account the fraction of ionizing radiation absorbed by dust grains and hence  not available
for ionizing hydrogen. The exact fraction depends mostly on the ionization parameter ($U$) of the NLR gas \cite[see][]{Netzer2013}.
The results shown below assume \Cf=0.05 and no dust absorption. This tends to overestimate the L(narrow \hb),
especially for $U>10^{-2}$.  

The calculations of L(\oiii) are more problematic. The line luminosity is sensitive to 
both the shape of the ionizing SED (known in this work) and the ionization parameter (not included in this work), 
and to a lesser extent on the
gas metallicity.
Two cases were considered: line luminosity which is scaled to \LION, and 
line luminosity which is scaled to L($E>35.1$ eV), where 35.1 eV is the ionization
potential of $O^+$. Both approximations are problematic because not all Lyman continuum photons can ionize $O^+$, 
and because of the possible
modification of the disk ionizing SED by soft X-ray radiation \cite[e.g.][]{Panda2019}. 
A reasonable approximation is L(\oiii)=0.0025(\Cf(NLR)/0.05)\LION\ but this may fail at extreme cases of very large BH mass or very small BH spin, where
L(E$>35.1$ eV) is very small. Because of this ambiguity, I do not provide calculated \kbol(\oiii) in this work.

Finally, L(\oi) is even more model dependent and no attempt was made to calculate \kbol(\oi).
Detailed photoionization calculations, and full discussion of these issues, including a comparison with a large sample
of type-II AGN, are given in \cite{Netzer2009b}.

\subsection{Bolometric correction factors}

The calculations of the various bolometric correction factors are presented in Fig.~\ref{fig:fig1} and Fig.~\ref{fig:fig2}. 
The bottom part of Fig.~\ref{fig:fig1} shows the correlation of \kbol(5100\AA) and \kbol(1400\AA) with \LAGN. There are four colors and point sizes that represent four BH masses:
\MBH=$10^7$~\Msun\ (tiny points), \MBH=$10^8$~\Msun\ (small points), \MBH=$10^9$~\Msun\ (large points) and \MBH=$10^{10}$\Msun\ (largest points). Each of the mass groups
 is divided into four groups of spin parameter:
-1 (lowest luminosity points),
0, 0.7, and 0.998 (highest luminosity points).
Each group of colored points shows a rise from bottom left to upper right following the increase in mass accretion rate,
 \Mdot. 
Since the behaviour of \kbol(3000\AA) is in between the other two, I chose not to show it in a separate panel.
It is important to emphasize that \kbol(3000\AA) refers to the disk SED. In many AGN
\cite[see e.g.][]{Mejia-Restrepo2016}, the so-called ``small blue bump'' (a blend of Balmer continuum and FeII line emission) gives an impression of a local continuum which may be as large as 20-50\% of the disk continuum. This must be taken into account when using this bolometric correction
factor to find \LAGN.

The top panels of 
Fig.~\ref{fig:fig1} and Fig.~\ref{fig:fig2}
show \kbol(2-10 keV) and \kbol(narrow \hb). 
 It is evident that the
scatter in all diagrams is very large because of the large range in accretion rate and BH spin. The scatter in
\kbol(2-10 keV) is somewhat smaller, mostly because this correction factor increases with \LAGN.

The calculations shown in Fig.~\ref{fig:fig1} do not provide a practical way to estimate \LAGN.
Therefore, I present in Fig.~\ref{fig:fig2} the calculated \kbol\ as a  function of
 measured  luminosities: L(1400\AA), L(5100\AA), L(2-10 keV) and L(narrow \hb). 
I also present eye-fitted log-log lines that go roughly through the points. The purpose is to properly
represent the case with $a=0.7$ ($\eta=0.104$) and hence the curves under-estimate \kbol\ for the highest \Mdot, maximum
BH spin cases. 

The simplified eye-fitted curves shown in Fig.~\ref{fig:fig2} are given by:
\begin{equation}
\kbol= c \times [L(observed)/10^{42}~erg/sec)]^d \,\, ,
\end{equation}
where L(observed) is the luminosity in question.                                                            
Table 1 presents recommended values for the constants $c$ and $d$. For the constant {\it c} in the expression
for \kbol(3000\AA) I give two approximations: one relative to the accretion disk continuum and one, in parentheses,
for a case where 
the Balmer continuum and FeII line blends are not removed prior to the estimate of \LAGN. 

There is no simple way to estimate the uncertainties in \kbol\ 
since the number of points shown in the various diagrams  
has nothing to do with the distribution of BH mass, spin, and accretion rate in the AGN population.
Moreover, the extreme values of \kbol\ depend on the various constraints imposed here on \Ledd, \LION/\LAGN\ and 
\LX/\LION. 
Fitting more
sophisticated functions to these points,  or using a $\chi^2$-type analysis is, therefore, not very meaningful. 
Perhaps a more meaningful way is to consider the extreme values of the various calculated \kbol. This shows
that, within the above constraints, the range   
is typically as large as the range in
the mass conversion efficiency $\eta$ (a factor of $\sim 8.4$) with some cases exceeding an order of magnitude. 
The range is considerably reduced at the
high luminosity end where other considerations, such as the fraction of ionizing radiation, limit the observable properties.

\begin{table}
\begin{center}
\caption{Fitting constants for  \kbol= $c \times [L(observed)/10^{42}~erg/sec)]^d$
 assuming ``mean'' disk inclination.
The number in parentheses is for cases where the Balmer continuum and the  FeII lines are included in L(3000\AA).
}
\begin{tabular}{lcc} 
$L(observed)$ & $c$   & $d$        \\
\hline
L(5100\AA)    &  40   & -0.2       \\
L(3000\AA)    &  25(19) & -0.2       \\
L(1400\AA)    &  7    & -0.1       \\
L(2-10 keV)   &  7    &  0.3       \\
L(narrow \hb) & 4580  & 0.18       \\
\end{tabular}
\end{center}
\end{table}

\begin{figure}
\centering
\includegraphics[width=8cm]{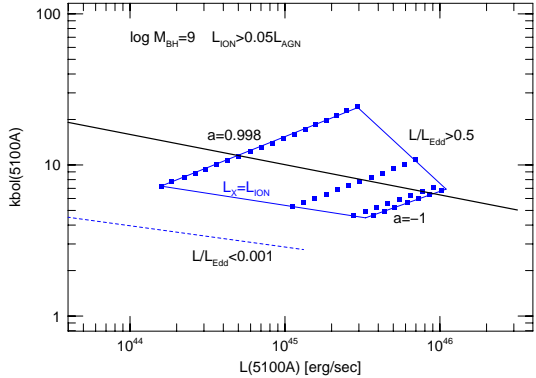}
 \caption{
Boundaries for \kbol(5100\AA) for the case of \MBH$=10^9$~\Msun. The boundaries set by the BH spin are marked by the spin parameter $a$.
The region on the right is not allowed since \Ledd$>0.5$ and the region below the lowest dashed line is where \Ledd$<0.001$.
The thick solid line is the approximation of eqn.~3.
}
\label{fig:fig3}
\end{figure}

Fig.~\ref{fig:fig3} provides more information about the range of \kbol(5100\AA) for the case of \MBH$=10^9$~\Msun.
The diagram shows the spin parameters considered earlier for various limits imposed by \LION\ and \LX. 
The solid lines show the boundaries of the allowed region for \LION$>0.05$\LAGN\ and \LX$<$\LION, 
the same as in Fig.~\ref{fig:fig1} and Fig.~\ref{fig:fig2}.
The boundaries are set by the smallest
($a=-1$) and largest ($a=0.998$) BH spins, and by the smallest and largest \Ledd\ which are consistent with these conditions. 
While the upper limit of \Ledd=0.5 is reached for all values of
{\it a}, the lowest Eddington ratio in this case (about 0.007) is much larger than the allowed lowest value of 0.001 (also
shown in the diagram). This was obtained by changing the condition on \LX\ to be
\LX$<0.5$\LAGN, which allows smaller values of \Ledd, in some cases as small as 0.001 (shown as
a dashed line).  
The thick solid line marks the approximation of eqn.~3, as in Fig.~\ref{fig:fig2}. 
For the default case where \LX$<$\LION, the uncertainty on \LAGN\ range from 
$\pm 0.05$  dex to $\pm 0.35$ dex, depending on L(5100\AA).
Clearly,  prior knowledge
of \MBH\ can significantly improve the determination of \LAGN.

Of the few earlier estimates of \kbol, those of \cite{Nemmen2010} and \cite{Runnoe2012a,Runnoe2012b} are most relevant to this work.
The \cite{Nemmen2010} paper made use of a library of accretion disk SEDs computed by Hubeny and collaborators. While these models are superior to the
disk models used here, the 
differences regarding \kbol\ are very small. 
The best fit approximation recommended by the authors is shown by a dashed black line in Fig.~\ref{fig:fig2}. This line is similar to 
the eye-fitted approximation of the present work
(note that Nemmen and Brotherton calculated \kbol(1450A) which is practically identical to \kbol(1400\AA) considered here).
The expression  obtained by \cite{Nemmen2010} is based on a $\chi^2$ method applied to a 
 library of AD models with various mass, spin and accretion rate. It depends on the availability of such models and not on population properties, and is hence not superior to 
 the eye-fitted curve used here.

\cite{Runnoe2012a} used a broken power-law SED, extending to 8 keV, to calculate a different set of bolometric correction factors. This is also shown in Fig.~\ref{fig:fig2}
by a dotted black line. Since the broken power-law continuum was chosen to correctly connect the optical and X-ray points, the resulting \kbol\ is not very different from 
the ones computed here. 
Fig.~\ref{fig:fig2} also shows the constant \kbol(5100\AA) and \kbol(1350\AA) used in the \cite{Shen2011} AGN catalog as adapted from the analysis in \cite{Richards2006}.

I also show in Fig.~\ref{fig:fig2} the widely used approximation of \cite{Marconi2004} which agrees with the AD approximation very well assuming \LX$<$\LION. For the alternative possibility  of 
\LX$<0.5$\LAGN\ (not shown here), the eye-fitted line
would move down by about 0.3 dex.
As shown by \cite{Vasudevan2007}, and also by \cite{Jin2012b}, the \cite{Marconi2004}
approximation gives too
small \kbol(2-10 keV) for NLS1s which is  
not surprising as such objects do not obey the basic relationship (eqn. 2) between L(2500\AA) and 
$L_{\nu (2 keV)}$ assumed here. 
The reason may be related to the type of ADs powering such sources (probably slim ADs
with large \Ledd). Because of this, \kbol(2-10 keV) estimated here should not be used to estimate \LAGN\ in NLS1s.  

Finally, the new calculations can be compared to those presented in \cite{Netzer2009b} where the
 recommended numbers  for \kbol(narrow \hb), assuming galactic-type extinction,
 is in the range 5600-8700 (shown in Fig.~\ref{fig:fig2}), for L(\oiii)/L(narrow \hb)$<12$. 
The new results are  
in reasonable agreement with this estimate 
especially given the uncertain amount of ionizing radiation absorbed by dust grains and the scatter in covering factor.
The reason for the large increase in \kbol(narrow \hb) at high L(narrow \hb) is 
the decrease in \LION/\LAGN\ at the high mass end.

\begin{figure}
\centering
\includegraphics[width=8cm]{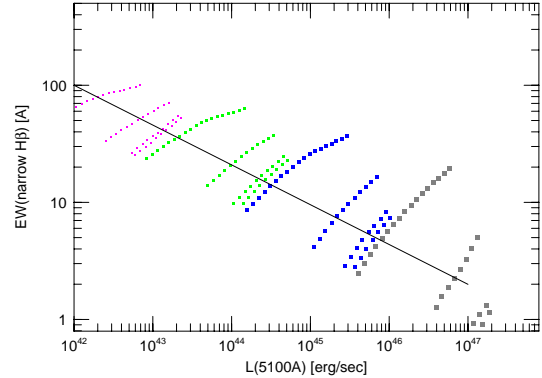}
 \caption{
Calculated EW(narrow \hb) for thin ADs assuming \Cf=0.05 and no absorption of ionizing radiation by dust grains inside
the NLR clouds. 
}
\label{fig:fig4}
\end{figure}

The new models can be used to calculate EW(narrow \hb) using the computed L(5100\AA) and L(narrow \hb). 
This is shown in Fig.~\ref{fig:fig4}. The numbers can be compared with observations of type-I AGN with little or no
dust reddening of the disk continuum. A simple eye-fitted curve gives:
EW(narrow \hb)=100(\Cf(NLR)/0.05)[L(5100\AA)/10$^{42}]^{-0.34}$.
As noted earlier, the calculations are not suitable for estimating the
\oiii\ luminosity and hence also not for estimating EW(\oiii).
 Some idea of the oxygen line luminosity and equivalent width can be obtained by noting that most 
type-II AGN fall in the range $2 < $L(narrow \oiii)/L(narrow \hb)$ < 10$ where high ionization AGN occupy the upper part
and LINERs the lower part of this range.

\section{Implications to AGN study}

The calculations presented here illustrate the big uncertainties associated with the estimate of \LAGN\ using bolometric correction factors based on various measured luminosities.
Several commonly used estimates  provide reasonable approximations to the mean population properties but can be wrong, by an order of
magnitude or more, for individual sources. BH mass estimates can significantly improve this situation (see Fig.~\ref{fig:fig3}). 
In addition, in low redshift, low luminosity type-I AGN, where the part of the SED with $L_{nu} \propto \nu^{1/3}$ is directly observed,
 \Mdot\ can be estimated from the luminosity of the long wavelength continuum
\cite[e.g.][]{Collin2002,LaorDavis2011,Netzer2014a}. In such cases, the  combination of \MBH\ and \Mdot\ limit the uncertainty to the unknown BH spin
and measuring optical, UV
or X-ray luminosity can reduce it further.

For type-I AGN, the unknown disk inclination is another source of uncertainty. All the calculations presented here are for an  inclination of 
$\sim 56$ degrees. For a torus
covering factor of 0.5, the mean expected inclination is smaller, corresponding to $\cos \theta \sim 0.7$. This 
would over-estimates the disk luminosity by a factor of $\sim 1.4$ and requires a 
corresponding decrease in \kbol(1400\AA), \kbol(3000\AA) and \kbol(5100\AA). For face-on ADs, the number is closer to 2.5. 

An interesting feature in Fig.~\ref{fig:fig2} is the convergence of \kbol(1400\AA) to a constant value of 2-3 at large AGN luminosity and large \MBH. 
The small range in \kbol\ is due to the combination of the small \LION/\LAGN\ typical of such sources, and the restrictions imposed here 
on \Ledd.
For the range of BH spin and \Ledd\ allowed here, 
\begin{equation}
\log \LION \simeq 12.9+0.72 \log L(1400\AA) \, .
\end{equation}
 For L(1400\AA)$>10^{47}$~\ergs, \LION\ is considerably smaller than \LAGN\ and its exact value, which is sensitive
to the exact value of $a$ and \MBH, does not influence much the derived \kbol.

An important potential source of uncertainty in \kbol(2-10 keV) is the unknown radiation pattern of the X-ray corona.
 In this work I assumed isotropic X-ray emission. This means that
\kbol(2-10 keV) is the same for type-I and type-II AGN. An X-ray radiation pattern similar to a thin disk pattern would result in 
a larger \kbol(2-10 keV) for type-II sources. There are other implications to the X-ray radiation pattern like those discussed in \cite{Netzer1987}.

Finally, Fig.~\ref{fig:fig2} and Fig.~\ref{fig:fig4} 
 suggest that EW(narrow \Hb), and the equivalent widths of other narrow emission lines, such as \OIII,
 are much smaller in high luminosity, large 
BH mass sources, provided the NLR covering factor is not increasing with luminosity.
There are indications that this is indeed observed in the most luminous AGN \cite[e.g.][]{Shemmer2004,Netzer2004}, although other explanations (an NLR size that
exceeds the size of the host galaxy) have been proposed.

\section{conclusions}
Optically thick geometrically thin ADs are likely to be the power-house of many AGN. 
The case of a disk which extends all the way to
the ISCO allows simple, straight-forward computations
of various bolometric correction factors that enable more robust estimates of \LAGN\ that are not sensitive to the details of the X-ray production
mechanism.
Such approximations are superior
to various constant bolometric correction factors used in the literature.

The new calculations provide simple approximations for \kbol(5100\AA), \kbol(3000\AA) and \kbol(1400\AA) for type-I AGN, and
 \kbol(2-10 keV) and \kbol(narrow \hb) for both types of AGN, excluding LINERs, over a large range of BH mass, BH spin
and BH accretion rate.
Their use can lead to a better evaluation of \LAGN\ and \Ledd\ in various situations of astrophysical interest. 
The paper 
demonstrates the large
inherited uncertainty in estimating \LAGN\ in cases where the BH spin and accretion rate are not known. The unknown inclination
of the disk in type-I AGN is an additional source of uncertainty.
Previous knowledge of the BH mass can reduce this uncertainty considerably. 

\section{Acknowledgments}

I am grateful to Benny Trakhtenbrot for useful comments and suggestions. 
I also thank the referee, Chris Done, for comments and suggestions that helped to clarify
the main points  of this work.



\end{document}